\documentclass[aps,prd,reprint,amsmath,amssymb,showpacs,superscriptaddress]{revtex4-1}

\usepackage{graphicx}
\usepackage{subfigure}
\usepackage{dcolumn}
\usepackage{bm}
\usepackage{slashed}
\usepackage{epstopdf}
\usepackage{mathrsfs}
\usepackage{color,xcolor}
\usepackage{appendix}

\usepackage{hyperref}
\newcommand{\td}{\ensuremath{\textrm{d}}}

\begin{document}

\title{QCD phase transition and equation of state of stellar strong interaction matter via Dyson-Schwinger equation approach}

\author{Zhan Bai}
\email{baizhan@pku.edu.cn}
\affiliation{Department of Physics and State Key Laboratory of Nuclear Physics and Technology, Peking University, Beijing 100871, China.}

\author{Yu-xin Liu}
\email{Corresponding author. yxliu@pku.edu.cn}
\affiliation{Department of Physics and State Key Laboratory of Nuclear Physics and Technology, Peking University, Beijing 100871, China.}
\affiliation{Collaborative Innovation Center of Quantum Matter, Beijing 100871, China.}
\affiliation{Center for High Energy Physics, Peking University, Beijing 100871, China.}


\date{\today}

\begin{abstract}
We study the phase structure and phase transition of cold dense QCD matter via the Dyson-Schwinger equation approach.
We take the rainbow approximation and the Gaussian-type gluon model.
In order to guarantee that the quark number density begins to appear at the nuclear liquid-gas phase transition chemical potential,
we propose a chemical potential dependent modification factor for the gluon model.
We find that for the iso-symmetric quark matter, the modification reduces the chemical potential of the phase coexistence region of the first--order phase transition.
We also implement the relativistic mean field theory to describe the hadron matter,
and make use of the Maxwell and Gibbs construction method to study the phase transition of $\beta$--equilibrium and charge neutral matter in compact stars.
The results show that the phase transition will not happen in case of the Gaussian--type gluon model without any modification.
The results also indicate that the upper boundary of the coexistence region should be larger than the current Nambu solution existing region.
We also calculate the mass-radius relation of the compact stars,
   and find that the hadron-quark phase transition happens at too high chemical potential so that the maximum mass of the compact star is hardly affected by the hadron-quark phase transition.
\end{abstract}

\maketitle

\section{Introduction}
\label{Sec:introduction}
The quantum-chromodynamics(QCD) is an asymptotic theory.
In vacuum, the quarks, which are the fundamental particles of the theory,
   are confined inside hadrons and cannot be detected directly.
At high temperature or/and high density, however,
   the interaction becomes weak, the quarks can the be released from hadrons.
The interaction responsible for the confinement also generates over $98\%$ mass of visible matter,
	and this effect is known as the dynamical chiral symmetry breaking (DCSB) (see, e.g., Refs.~\cite{Roberts:1990PRD,LChang:2007,Williams:2007PLB,Fischer:2009EPJC,Fischer:2011PRD,Wang:2012PRD,Mitter:2014PRD,Braun:2014PRD}).
The complexity of QCD leads to an abundant of phase structures.
At zero chemical potential and finite temperature,
   it is believed that the transition from hadron matter to quark matter is an crossover~\cite{Aoki:2006Nature,Aoki:2009JHEP,Borsanyi:2010JHEP,Bazavov:2012PRD,Bhattacharya:2014PRL,Bazavov:2014PRD,Qin:2011PRL,Fischer:2011PLB,Fischer:2013PLB,Fischer:2014PRD,Eichmann:2016PRD,Gao:2016PRDa,Gao:2016PRDb,Gao:2016PRDc,Fischer:2019PPNP,Fu:2020PRD,Gao:2020PRD,Dupuis:2020arXiv}.
At low temperature and high chemical potential,
    although there are arguments that the transition may also be a crossover~\cite{Schfer:1999PRL,Kitazawa:2002PTP,Hatsuda:2006PRL,Zhang:2009PRD,Kojo:2016EPJA},
  most model calculations show that this phase transition is of first--order~\cite{Asakawa:1989NPA,Stephanov:1998PRL,Schaefer:2005NPA,Herbst:2011PLB,Qin:2011PRL,Gao:2016PRDa,Gao:2016PRDb,Otto:2020PRD}.
Therefore, there should exist a critical end point (CEP) on the $T-\mu$ plane,
	which connects the crossover region with the first--order phase transition,
	and the searching for the CEP is a hot topic in the study of QCD phase structure~\cite{Halasz:1998PRD,Berges:1999NPB,Qin:2011PRL,Fischer:2011PLB,Fischer:2013PLB,Fischer:2014PRD,Eichmann:2016PRD,Gao:2016PRDa,Gao:2016PRDb,Gao:2016PRDc,Fischer:2019PPNP,Fu:2020PRD,Gao:2020PRD,Dupuis:2020arXiv}.
At low temperature and high chemical potential, the phase structure is even more complicated.
At baryon chemical potential $\mu_{B}\sim 923\;$MeV, there is a first--order liquid-gas phase transition where the nuclear matter emerges from vacuum~\cite{Fukushima:2013PPNP,Langelage:2014JHEP,Weyrich:2015PRC,Pochodazalla:1995PRL,Chomaz:2001NPA,DAgstino:2005NPA,Elliot:2013PRC}.
After that, there is the phase transition from nuclear matter to quark matter.
It is also possible that the hadron-quark phase transition will lead to color-superconducting quark phase
(see Ref.~\cite{Alford:2008RMP} for review and some recent works in Ref.~\cite{Mueller:2016,Flores:2017PRC,Roupas:2020,Bogadi:2020PRD,Rocha:2020IJMPD}).

At sufficiently high temperature or/and large chemical potential,
   since the interaction is weak, the perturbative QCD approach can provide reliable results on the property of QCD matter~\cite{Kurkela:2010PRD,Kurkela:2016PRL}.
However, in the phase transition region, the interaction is strong, and perturbative approach becomes invalid.
One must then implement non-perturbative approaches to study the phase structure.
There are phenomenological models such as NJL model,
	  the quark meson model and their Polyakov loop improved versions~\cite{Buballa:2005PR,Fukushima:2004PLB,Megias:2006PRD,Ratti:2006PRD,Schaefer:2007PRD,Fu:2007PRD,Skokov:2010PRC,Herbst:2011PLB}.
Also, there are first principle methods such as lattice QCD~\cite{Aoki:2006Nature,Aoki:2009JHEP,Borsanyi:2010JHEP,Bazavov:2012PRD,Bhattacharya:2014PRL,Bazavov:2014PRD,Ding:2019PRL,Ding:2020},
functional renormalization group theory~\cite{Braun:2009PRL,Pawlowski:2014NPA,Mitter:2014PRD,Braun:2014PRD,Fu:2020PRD,Gao:2020PRD,Dupuis:2020arXiv,Braun:2020PRDa,Braun:2020PRDb}
and Dyson-Schwinger equation approach (for some reviews, see, e.g., Refs.~\cite{Roberts:2000PPNP,Roberts:2012CTP,Fischer:2019PPNP}).

It is well known that lattice QCD is principally only valid at zero chemical potential, since it endures the notorious ``sign problem" at finite chemical potential.
Although there are extrapolation methods such as the Taylor expansion~\cite{Allton:2002PRD,Gavai:2003PRD,Allton:2003PRD,Kaczmarek:2011PRD,Bazavov:2019PLB},
the imaginary chemical potential~\cite{Forcrand:2002NPB,Elia:2003PRD,Elia:2007PRD,Forcrand:2010PRL,Laermann:2013JPCS,Bonati:2014PRD,Philipsen:2016PRD},
and other approach (e.g., Ref~\cite{Liu:2011PRD}), the lattice QCD can still only deal with small chemical potential $\mu/T<2$.
Therefore, the theoretical study of the phase transition at zero temperature and finite chemical potential is not as clear as that in zero chemical potential.

The problem also exists on the experimental side.
There are colliders such as RHIC, FAIR, NICA and HIAF which aim at studying the structure of hot or warm dense QCD matter,
	  but terrestrial experiments are now not able to create cold dense QCD matter.
The way to study the possible phase transition at large chemical potential has then to resort the astronomical observations.
There have been several very heavy compact stars with a mass over $\sim 2M_{\odot}$ discovered~\cite{Demorest:2010Nature,Antoniadis:2013Science,Fonseca:2016APJ,Arzoumanian:2018APJS,Linares:2018APJ,Cromartie:2019NA},
and the detection of gravitational wave from binary neutron star merger also provide constraints on the equation of state (EOS) and hence the possible phase transition inside compact stars~\cite{Abbott:2017APJL,Margalit:2017APJL,Shibata:2017PRD,Abbott:2018PRL,Ruiz:2018PRD,Annala:2018PRL,Rezzolla:2018APJL,Shibata:2019PRD,Abbott:2020APJ,Annala:2020NP}.
Still, the detail of the phase transition is hidden inside compact stars.

It has been known that the Dyson-Schwinger (DS) equations approach (see, e.g., Refs.~\cite{Roberts:1994PPNP,Roberts:2000PPNP,Roberts:2012CTP,Alkofer:2001PR,Fischer:2006JPG}) is a continuous field theory of QCD,
which can simultaneously deal with the confinement and dynamical chiral symmetry breaking,
   and is available on the entire $T-\mu$ plane as well as zero current quark mass limit.
   The DS equations include in fact an infinite number of coupled integral equations, and in order to solve them,
	one must take truncations~\cite{Ball:1980PRD,Fischer:2007PRD,Chang:2011PRL,Eichmann:2016PPNP,Williams:2016PRD,Aguilar:2018PRD,Tang:2019PRD}
	and model the dressed gluon propagator~\cite{Munczek:1983PRD,Maris:1999PRC,Roberts:2011PRC,Qin:2011PRC}.
The DS equation approach has been taken to study the phase transition and the CEP~\cite{Qin:2011PRL,Fischer:2011PLB,Fischer:2013PLB,Gao:2016PRDa,Gao:2016PRDb,Gao:2016PRDc,Fischer:2014PRD,Reinosa:2015PRD,Eichmann:2016PRD,Reinosa:2017PRD,Maelger:2018PRDa,Maelger:2018PRDb,Gunkel:2019EPJA,Isserstedt:2019PRD,Fischer:2019PPNP,Gao:2020PRD,Maelger:2020PRD},
	and also been used to study the cold dense matter~\cite{Chen:2008PRD,Chen:2011PRD,Chen:2012PRD,Chen:2015PRD,Xu:2015IJMPA,Chen:2016EPJA}.

However, in spite of the fact that the DS equation approach recovers successfully many properties of the DCSB phase,
	it is still far from being able to describe the real-world hadron matter.
For example, one of the most important properties of hadron matter is the emergence of matter from vacuum,
    which corresponds to the first--order nuclear liquid--gas phase transition.
In DS equation calculations, it has been shown that the quark number density and quark condensate remains the same as in vacuum up to a critical chemical potential~\cite{Muller:2016arXiv,Fischer:2019PPNP,Gunkel:2020JPCS,Chang:2007PLB},
   but this critical chemical potential is not in accordance with the liquid--gas phase transition chemical potential.
Therefore, in this paper, we propose a chemical potential dependent modification on the coupling strength of the model in the approach
   to describe the phase transition in the cold dense matter,
   and construct the EOS of the hybrid star matter. The mass-radius relation of the highly massive compact star can be described well.

The remainder of this paper is organized as follow.
After this introduction, in Sec.~\ref{Sec::DSE}, we reiterate briefly the DS equation approach at zero temperature and finite chemical potential,
	  and propose the chemical potential dependent modification on the model.
In Sec.~\ref{Sec::result}, we present the numerical results with the DS equation being solved,
   including the phase transition of iso-symmetric matter as well as the $\beta$--equilibrium and charge neutral cold dense matter.
   We also calculate the EOS and the mass-radius relation of the compact star.
In Sec.~\ref{Sec::summary}, we give a summary and some remarks.

\section{Dyson-Schwinger equation approach}\label{Sec::DSE}

\subsection{Gap Equation at Zero Temperature and Finite Chemical Potential}

In this section, we describe briefly the DS equation approach at zero temperature and finite chemical potential.

The starting point of the DS equation approach is the gap equation:
\begin{equation}
S(p;\mu)^{-1}=Z_{2}^{} [i\boldsymbol{\gamma} \cdot  \boldsymbol{p} + i\gamma_{4}^{} (p_{4}^{} + i\mu)+m_{q}^{}]+\Sigma(p;\mu),
\end{equation}
where $S(p;\mu)$ is the quark propagator, $\Sigma(p;\mu)$ is the renormalized self-energy of the quark:
\begin{equation}
\begin{split}
\Sigma(p;\mu)=&\; Z_{1}^{} \int^{\Lambda}\frac{\textrm{d}^{4} q}{(2\pi)^{4}}
          g^{2}(\mu)D_{\rho\sigma}^{} (p-q;\mu)\\
		&\times\frac{\lambda^{a}}{2} \gamma_{\rho} S(q;\mu) \Gamma_{\sigma}^{a}(q,p;\mu),
\end{split}
\end{equation}
where $\int^{\Lambda}$ is the translationally regularized integral,
$\Lambda$ is the regularization mass-scale.
$g(\mu)$ is the strength of the coupling, $D_{\rho\sigma}^{}$ is the dressed gluon propagator,
$\Gamma_{\sigma}^{a}$ is the dressed quark-gluon vertex,
$\lambda^{a}$ is the Gell-Mann matrix, and $m_{q}^{}$ is the current mass of the quark.
$Z_{1,2}^{}$ are the renormalization constants.
In this paper, we take a model that ultraviolet integration is finite,
   so we can move the renormalization point to infinity and take $Z_{1,2}^{}=1$.

At finite chemical potential,
the quark propagator can be decomposed according to the Lorentz structure as:
\begin{equation}
\begin{split}
S(p;\mu)^{-1}=&\; i\boldsymbol{\gamma} \cdot  \boldsymbol{p} A(p^{2}, p \, u, \mu^{2})
+ B(p^{2},p\, u, \mu^{2})\\
		&+i\gamma_{4}^{} (p_{4} + i\mu) C(p^{2},p\, u, \mu^{2}) \, ,
\end{split}
\end{equation}
with $u=(\boldsymbol{0},i\mu)$.
A complete decomposition should include another term proportional to $\sigma_{\mu\nu}$,
  but this term contributes little, and is usually omitted~\cite{Roberts:2000PPNP,Fischer:2019PPNP}.

At zero chemical potential, a commonly used ansatz for the dressed gluon propagator and the dressed quark-gluon interaction vertex is:
\begin{equation}
Z_{1}^{} g^{2}  D_{\rho\sigma}^{}(p-q)\Gamma_{\sigma}^{a}(q,p)
		=\mathcal{G}((p-q)^{2} )D_{\rho\sigma}^{\textrm{free}}(p-q) \frac{\lambda^a}{2}
             \Gamma_{\sigma}^{}(q,p) \, ,
\end{equation}
where
\begin{equation}
D_{\rho\sigma}^{\textrm{free}}(k\equiv p-q)=\frac{1}{k^{2}} \Big( \delta_{\rho\sigma}^{} - \frac{k_{\rho}^{} k_{\sigma}^{}}{k^{2}} \Big) \, ,
\end{equation}
$\mathcal{G}(k^{2})$ is the effective interaction to be introduced in the model,
and $\Gamma_{\sigma}^{}$ is the quark-gluon vertex.
In this paper, we adopt the rainbow approximation which is the leading-order term in a symmetry preserving approximation scheme~\cite{Munczek:1995PRD,Bender:1996PLB},
\begin{equation}
\Gamma_{\sigma}^{}(q,p) = \gamma_{\sigma}^{} \, .
\end{equation}

For the interaction part, we adopt the Gaussian type effective interaction :
\begin{equation}\label{eqn:Gauss}
\frac{\mathcal{G}(k^{2})}{k^{2}}=\frac{4\pi^{2} D}{\omega^{6}}k^{2} \textrm{e}^{-k^{2}/\omega^{2}} \, ,
\end{equation}
where $D$ and $\omega$ are the parameters of the model.
In this paper we take $\omega=0.5\,$GeV and $D=1.0\,\textrm{ GeV}^2$ which are determined with the pion mass $m_{\pi}=0.14\;$GeV and decay constant $f_{\pi}=0.091\;$GeV with $m_{q}=5\;$MeV for $u$ and $d$ quark~\cite{Chang:2009PRL}.
And the current mass of the strange quark is $m_{q}=115\;$MeV by fitting the kaon mass $m_{K} =0.492\;$GeV~\cite{Alkofer:2002PRDb}.

The DS equation has multiple solutions.
In chiral limit (zero current quark mass) and zero chemical potential, there is a solution with $B(p)\equiv 0$,
   and a solution with $B(p)>0$, which are called Wigner solution, Nambu solution, respectively.
The Wigner solution corresponds to the dynamical chiral symmetry (DCS) phase, where the quarks are massless.
The Nambu solution corresponds to dynamical chiral symmetry breaking (DCSB) phase, since the mass function $M(p)=B(p)/A(p)$ acquires a non-zero value.
Although there are hints that there might exist a quarkyonic phase where the quark is confined but the chiral symmetry is preserving,
it is usually believed that the DCSB (Nambu) solution corresponds to the confined hadron phase,
   and the DCS (Wigner) solution corresponds to the deconfined quark phase.

At zero temperature and small quark chemical potential,
   the DCSB phase should stay in the vacuum ground state until the chemical potential is larger than some critical value.
This property is known as ``Silver--Blaze'' property of QCD.
At baryon chemical potential $\mu_{B}^{}=923\;$MeV, which is the proton mass minus the binding energy,
there will be a liquid-gas phase transition where nucleons begin to emerge from vacuum,
and the Silver--Blaze property will be broken beyond this chemical potential.

\subsection{Quark Number Density and Silver--Blaze Property}

After solving the DS equation at some quark chemical potential,
the number density of the quarks can be obtained through~\cite{Chen:2008PRD} :
\begin{equation}
n_{q}^{}(\mu) = 6\int\frac{\textrm{d}^{3}p}{(2\pi)^{3}} f_{q}^{}(|\boldsymbol{p}|;\mu),
\end{equation}
where $f_{q}^{}$ is the distribution function and reads
\begin{equation}\label{eqn:distribution}
f_{q}^{}(|\boldsymbol{p}|;\mu) = \frac{1}{4\pi}\int_{-\infty}^{\infty}\textrm{d}p_{4}^{} \textrm{tr}_{\textrm{D}}^{}[-\gamma_{4}^{} S_{q}^{}(p;\mu)] \, ,
\end{equation}
where the trace is for the spinor index.

The integration in Eq.~(\ref{eqn:distribution}) can be converted to a contour integral on the complex plane of $\tilde{p}_4$:
\begin{equation}\label{eqn:contour}
\begin{split}
f_{q}^{}(|\boldsymbol{p}|;\mu) =& \; \frac{1}{\pi}\int_{-\infty}^{\infty}\textrm{d}p_{4}^{} \frac{i\tilde{p}_{4}C(|\vec{\,p}|^2,\tilde{p}_{4}^2)}{\mathcal{M}} \\
		=&\frac{1}{\pi}\int_{-\infty+i\mu}^{\infty+i\mu}\td \tilde{p}_{4} \frac{i\tilde{p}_{4}C(|\vec{\,p}|^2,\tilde{p}_{4}^2)}{\mathcal{M}} \\
		=&\frac{1}{\pi}\Bigg(\oint_{(-\infty+i\mu)\rightarrow(\infty+i\mu)\rightarrow(\infty)\rightarrow(-\infty)\rightarrow(-\infty+i\mu)}\\
&-\!\int_{\infty}^{-\infty} \!\!- \!\!\int_{-\infty}^{-\infty+i\mu} \!\! - \!\! \int_{\infty+i\mu}^{\infty}\! \Bigg)\!\frac{i\tilde{p}_{4}C(|\vec{\,p}|^2,\tilde{p}_{4}^2)\td\tilde{p}_{4}}{\mathcal{M}} , \\
\end{split}
\end{equation}
where
\begin{equation}\label{eqn:denominator}
\mathcal{M}=\vec{\,p}^{2}A^{2}(|\vec{\,p}|^2,\tilde{p}_{4}^2)+\tilde{p}_{4}^{2}C^2(|\vec{\,p}|^2,\tilde{p}_{4}^2) + B^2(|\vec{\,p}|^2,\tilde{p}_{4}^2).
\end{equation}

In the last line of Eq.~(\ref{eqn:contour}), the integral $\int_{\infty}^{-\infty}$ is zero since the kernel is an odd function,
   and the last two integrals are also zero because the denominator $\mathcal{M}$ diverges at infinity.
Therefore, the value of distribution function $f_{q}$ is determined by the poles inside the contour.

\begin{figure}[!htb]
\includegraphics[width=0.45\textwidth]{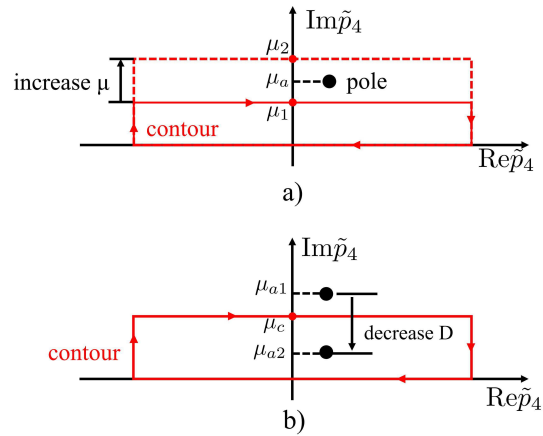}
\vspace*{-4mm}
\caption{Schematic feature of the contour and the singularities.
	{\it upper panel}: By increasing the chemical potential, the contour may include the pole, and the number density becomes nonzero.
	{\it lower panel}: For a fixed chemical potential, by decreasing the coupling constant $D$, the imagin part of the pole becomes small and moves inside the contour, and the number density becomes nonzero.
}
\label{fig:singularity}
\end{figure}

The pole corresponds to the zero in the denominator $\mathcal{M}$:
\begin{equation}\label{eqn:pole}
\tilde{p}_{4}=\pm\frac{i\sqrt{\vec{p}^2A^2+B^2}}{C}.
\end{equation}
A schematic figure of the pole and the contour is shown in the upper panel of Fig.~\ref{fig:singularity}.
For Nambu solution, since $B$ is non-zero, we can see from Eq.~(\ref{eqn:pole}) that the imaginary part of the first pole, $\mu_{a}$, is finite for arbitrary value of $\vec{p}^2$,
and for the chemical potential $\mu<\mu_{a}$, the integral in Eq.~(\ref{eqn:contour}) is zero since there is no singularities inside the contour,
	and the quark number density remains zero.
When $\mu > \mu_{a}$, the contour begins to include the pole, and the number density becomes nonzero.
This is a manifestation of the Silver--Blaze property.

\subsection{Modification for the Gluon Model}

In DS equation calculations, the Silver--Blaze property is maintained.
The condensation and number density of the Nambu solution remains the same as in vacuum for small quark chemical potential (see, e.g. Ref.~\cite{Chen:2008PRD,Chang:2007PLB,Huang:2020PRD}).
However, the critical chemical potential where quark number density becomes non-zero relies on the vertex and the gluon model,
	and is usually not in accordance with the nuclear liquid-gas phase transition chemical potential.

We notice that, by altering the coupling constant $D$ in Eq.~(\ref{eqn:Gauss}), the critical chemical potential can be changed.
The schematic characteristic of the procedure is shown in the lower panel of Fig.~\ref{fig:singularity}.
For a fixed chemical potential, the pole may lie outside the contour of Eq.~(\ref{eqn:contour}), and the number density is zero.
However, by decreasing the coupling constant $D$, the imaginary part of the pole should also decrease.
When the pole moves into the contour, the number density will becomes nonzero.

In order to correctly recover the liquid-gas phase transition chemical potential,
   we are going to add a chemical potential dependence to the coupling $D$ by multiplying a modification function $h(\mu)$ to the coupling constant $D$ defined in Eq.(\ref{eqn:Gauss}), so as to have
\begin{equation}
D(\mu)=D(\mu=0)h(\mu),
\end{equation}
where $D(\mu=0)=1.0\;\textrm{GeV}^{2}$.

There are three constraints on the modification function $h(\mu)$:
First, we should have $h(\mu=0)=1$ since the vacuum value is fixed by meson properties.
Second, $h(\mu=\infty)=0$ to take into account the asymptotic freedom at large chemical potential.
And we also requires that $h(\mu)$ guarantees that the critical chemical potential is $\mu_{B,c}=923\,$MeV which is in accordance with the emergence of nuclear matter.

Previously, in the study of compact star matter, an exponentially damping function is proposed~\cite{Chen:2011PRD,Chen:2012PRD,Chen:2015PRD,Chen:2016EPJA,Bai:2018PRD} as:
\begin{equation}\label{eqn:Damping_ch}
h_{\alpha}(\mu_{q})=\exp(-\alpha\mu_{q}^2/\omega^2),
\end{equation}
where $\alpha$ is a parameter and $\omega$ is the same as in the original gluon model.
This function is a monotonically decreasing function.
In general, a larger coupling constant $D$ corresponds to a larger quark mass function $M(p)$,
   and a larger critical chemical potential.
Meanwhile it reduces the value $\mu_{B,c}^{}$ for the phase transition to happen~\cite{Bai:2018PRD}.

However, for some vertex and gluon propagator, the $\mu_{B,c}^{}$ might be less than $923\,$MeV (see, e.g.Ref.~\cite{Huang:2020PRD}),
	which indicates that a non-monotonic modification function is required.
In this work, we introduce another type of modification function:
\begin{equation}\label{eqn:Damping_bz}
h_{\beta}(\mu_{q})=\left(1+\frac{\mu_{q}^2}{\mu_{q,c}^2}\right)\exp(-\beta\mu_{q}^2/\omega^2),
\end{equation}
The value of the parameters $\alpha$ and $\beta$ are fixed by requiring $\mu_{B,c}^{}=923\,$MeV (i.e., $\mu_{q,c}^{}=307.6\,$MeV).
In the following, we will denote the modification function in Eq.(\ref{eqn:Damping_ch}), in Eq.(\ref{eqn:Damping_bz}) as $\alpha$ type, $\beta$ type, respectively.

\section{Results and Discussions}\label{Sec::result}

\subsection{Determination of parameter}

The Nambu solution of the quark's DS equation has a relatively large mass function.
Therefore at small chemical potential, the number density of quark in Nambu solution should be zero.
The mass function is related to the coupling constant $D$,
    so for a fixed chemical potential, the quark number density relies definitely on the coupling strength $D$.

For $D/D_{0}=1$, where $D_{0}=1.0\,\textrm{GeV}^2$ is the coupling strength in vacuum and fixed with the meson properties,
the number density should be zero~\cite{Chen:2008PRD}, which satisfies the Silver-Blaze property.
However, because of numerical error, the calculated number density is non-zero and numerically unstable.
Therefore, it is hard to determine the critical coupling strength via the number density directly.

\begin{figure}[!htb]
\vspace*{-3mm}
\includegraphics[width=0.45\textwidth]{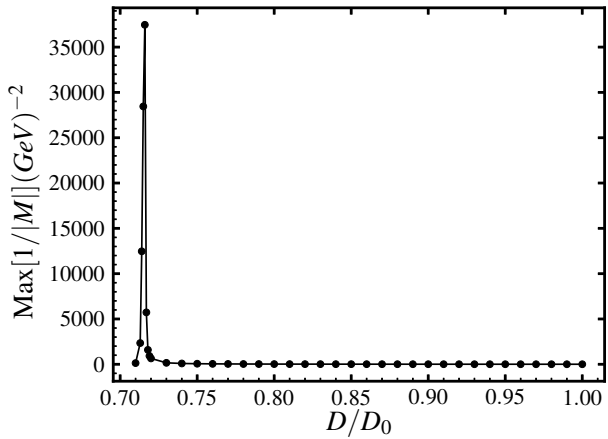}
\vspace*{-4mm}
\caption{Calculated coupling strength $D$ dependence of the denominator $1/|\mathcal{M}|$ defined in Eq.~(\ref{eqn:denominator}), with the critical chemical potential $\mu_{q,c}=307.6\;$MeV, and dressed gluon propagator in the Gaussian--type with $D = D_{0}=1.0\;\textrm{GeV}^{2}$.
}
\label{fig:D_critical}
\end{figure}

Another way to determine the critical coupling strength is by searching for the mass pole.
As we have stated in the last section, the emergence of the quark number density corresponds to a mass pole in the propagator.
The denominator $\mathcal{M}$, as defined in Eq.~(\ref{eqn:denominator}), is a function of $|\vec{\,p}|$ and $p_{4}^{}$, for a fixed value of $\mu=\mu_{q,c}$.
When the pole enters the contour of the integral in Eq.~(\ref{eqn:contour}), as illustrated in the lower panel of Fig.~\ref{fig:singularity},
	 there should be a zero $\mathcal{M}$ for a certain value of $|\vec{\,p}|$ and $p_{4}$,
	 or equivalently, the maximum value of $1/|\mathcal{M}|$ should be divergent.
The obtained maximum value of the $1/|\mathcal{M}|$ as a function of coupling constant is shown in Fig.~\ref{fig:D_critical}.
From the figure, it is apparent that $1/|\mathcal{M}|$ has a pole at $D_{c}=0.716D_{0}$, where the quark number density begins to appear.

\begin{figure}[!htb]
\includegraphics[width=0.45\textwidth]{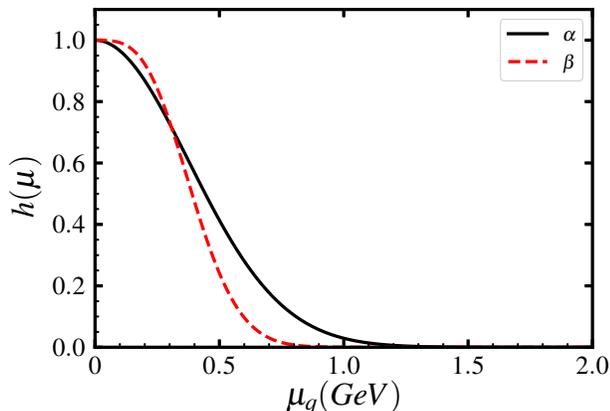}
\vspace*{-4mm}
\caption{Calculated quark chemical potential dependence of the modification coefficient $h$.
The black solid line corresponds to the $\alpha$-type modification function defined in Eq.(~\ref{eqn:Damping_ch}),
and the red dashed line corresponds to the $\beta$-type modification function defined in Eq.(~\ref{eqn:Damping_bz}). }
\label{fig:Damping_Factor}
\end{figure}

The parameter $\alpha$ and $\beta$ in Eq.(\ref{eqn:Damping_ch}) and Eq.(\ref{eqn:Damping_bz}) is fixed by requiring that $h(\mu_{q,c})\times D_{0}=D_{c}$,
    and we have $\alpha=0.883$ and $\beta=2.714$.
The calculated variation behavior of the modification coefficient $h$ with respect to the quark chemical potential $\mu_{q}$ (i.e., $\mu_{B}^{}/3$) is shown in Fig.~\ref{fig:Damping_Factor}.
The black solid line corresponds to the result with the $\alpha$--type modification function defined in Eq.(~\ref{eqn:Damping_ch})
and the red dashed line corresponds to the $\beta$-type modification function defined in Eq.(~\ref{eqn:Damping_bz}).
Both of the two presently obtained modification functions are monotonically decreasing.
However, at small chemical potential,
   the modification $h_{\alpha}(\mu)$ is smaller than the $h_{\beta}(\mu)$,
while at large chemical potential  $h_{\beta}(\mu)$ decreases to almost zero more rapidly.
The two lines cross at $\mu_{q}=307.6\;$MeV,
    which is required by our assumption.
After fixing the parameters, we can solve the DS equations with the Gaussian-type dressed gluon model defined in Eq.(\ref{eqn:Gauss}) without any modification,
	  with the $\alpha$-type modification and the $\beta$-type modification, respectively.

\subsection{Phase Transition of the Iso-symmetric Matter and the Coexistence Region}\label{sec::isosymmetric}

Because of the asymptotic nature of QCD, there must exist a deconfined phase transition at high density (chemical potential).
Regardless of the possible existence of quarkyonic phase, this phase transition is also the chiral phase transition from DCSB phase to DCS phase.

The order parameter of the chiral phase transition reads usually the condensate of quark,
    which is the trace of the quark propagator.
The chiral susceptibility, which is the derivative of the quark condensate with respect to the current quark mass,
    is taken as the criteria of the phase transition.

The chiral susceptibility is related to the stability of the system~\cite{Qin:2011PRL,Gao:2016PRDb}.
If the chiral susceptibility is positive, the system is stable or meta-stable,
and the system is unstable if the chiral susceptibility is negative.

In vacuum, the chiral susceptibility of the Nambu solution, which corresponds to the DCSB phase, is positive,
while the chiral susceptibility of the Wigner solution, which corresponds to the DCS phase, is negative(see, e.g. Ref.~\cite{Qin:2011PRL}).
This means that the DCS quark matter is unstable in vacuum, and the system should be described by the Nambu solution.
When the chemical potential gets large enough, the Nambu solution will disappear,
and the chiral susceptibility of the Wigner solution is positive,
   this means that the quarks gets deconfined from hadrons and the matter should be in quark phase.

In chiral limit,
   the quark condensate and the chiral susceptibility is well defined.
In case of nonzero current quark mass, however, both the quark condensate and the chiral susceptibility with the original definition in tracing the propagator diverges drastically.
There have been several subtract schemes to eliminate the divergence (see, e.g., Refs.~\cite{LChang:2007,Williams:2007PLB,Qin:2011PRL,Bazavov:2012PRD,Gao:2016PRDb,Fischer:2019PPNP,Braun:2020PRDb},
	but it is quite convenient to simply make use of the Lorentz scalar part of the quark propagator at zero momentum,
{\it i.e.,} taking $B(p=0)$ as a representative of the order parameter,
and  $\chi_{m}\equiv \frac{\td B(p=0)}{\td m_{0}}$ as chiral susceptibility,
as has been done in many literatures.
Specifically, it has been shown that, for the phase transition at zero chemical potential and high temperature, the value of the pseudo-critical temperature would be slightly different by using this criterion~\cite{Gao:2016PRDb}.
At zero temperature and high chemical potential, however,
the chiral susceptibility will diverge at the phase boundary, and the phase transition chemical potential determined with this definition is not different from those fixed via other forms of the defination.

\begin{figure}[!htb]
\vspace*{-3mm}
\includegraphics[width=0.45\textwidth]{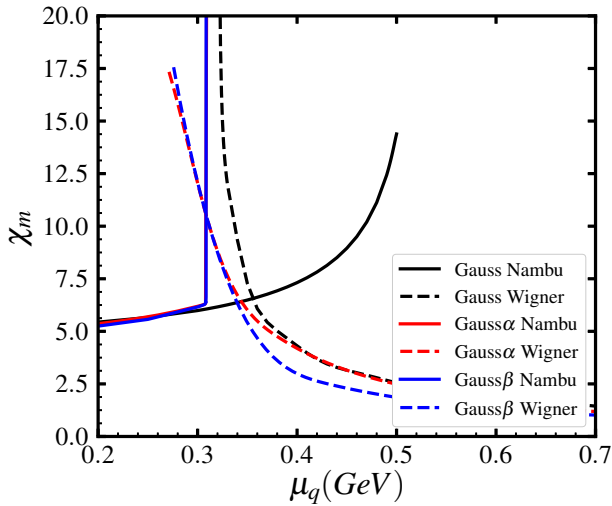}
\vspace*{-4mm}
\caption{(color online) Calculated chiral susceptibility with different gluon models.
The black lines correspond to the results in the Gaussian-type model without modification,
the red lines stand for the results implementing the $\alpha$-type modification function defined in Eq.(\ref{eqn:Damping_ch}),
and the blue lines represent the results using the $\beta$-type modification function defined in Eq.(\ref{eqn:Damping_bz}).
The solid lines correspond to that of the Nambu solution, and the dashed lines correspond to that of the Wigner solution.}
\label{fig:Mu_dBdm_ud}
\end{figure}

In Fig.~\ref{fig:Mu_dBdm_ud}, we show the calculated chiral susceptibilities of both the Nambu and the Wigner solutions with the different gluon models,
   and we have neglected the solution where the chiral susceptibility is negative.
As we can see from the figure, for all the three models,
   there exists a (quark) chemical potential region where both the Nambu solution and the Wigner solution have positive chiral susceptibility,
   and this region is usually referred to as the ``phase coexistence region",
   where the DCSB hadron matter and the DCS quark matter appear simultaneously.

In details, in case of the Gaussian-type gluon model without chemical potential dependence, the coexistence region is $\mu_{q} \in [0.322, 0.500]\,$GeV.
In case of the Gaussian-type gluon model with the $\alpha$-type modification, the coexistence region is $\mu_{q} \in [0.271 , 0.309]\,$GeV,
    and for the gluon model with the $\beta$-type modification, the coexistence region is $\mu_{q} \in [0.276, 0.308]\,$GeV.
Both the upper and lower boundaries of the coexistence region are smaller than the corresponding one in case of without the modification,
     and the coexistence region gets narrower.
This is because our modification coefficient generally reduces the interaction strength,
     and it will be easier for the quark to be released from hadrons.

\subsection{$\beta$--equilibrium Quark Matter}

In the last subsection, we have shown the calculated phase boundary and phase transition for the iso-symmetric matter,
{\it i.e.,} the chemical potentials of the $u$ and $d$ quarks are the same.
However, it is now impossible for us to get high density matter in terrestrial experiment.
The most common way to study the phase structure and the phase transition in the high density region resorts then studying the matter in compact stars.
The compact star matter is usually in $\beta$--equilibrium and charge neutral.
Therefore, we are now going to study the $\beta$--equilibrium and charge neutral quark matter.

To study of the property of compact star matter composed of both hadrons and quarks (hybrid matter),
it is presently common to implementing the theoretical approach of hadron matter and that of quark model separately to study hadron, quark phase, respectively,
and combine them with construction schemes.

In this section, we calculate the property of quark matter by solving the DS equation of quark,
and only keep the Wigner solution, since the Nambu solution should corresponds to the hadron phase.
For the hadron phase, however, we take the relativistic mean field (RMF) theory with TW-99 parameterization~\cite{Typel:1999NPA}, which is described in the appendix.

\begin{figure}[!htb]
\vspace*{-3mm}
\includegraphics[width=0.45\textwidth]{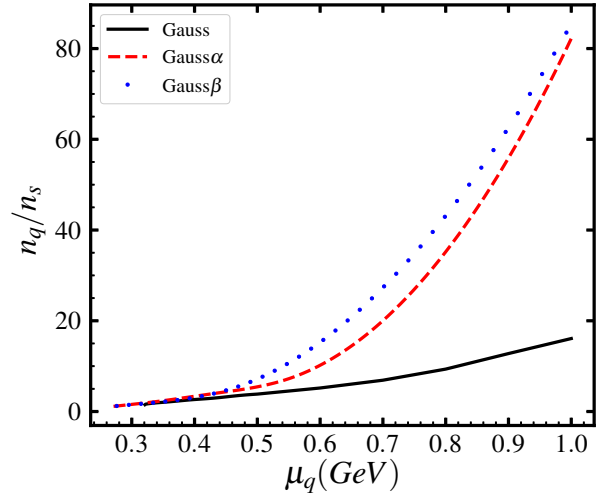}
\vspace*{-4mm}
\caption{(color online) Calculated number density (in unit of the saturation nuclear matter density $n_{S}$) of the $u$ and $d$ quarks as a function of quark chemical potential.
	The black solid line corresponds to the result using the Gaussian-type gluon model without modification,
	the red dashed line stands for the result by implementing the Gaussian-type gluon model with the $\alpha$--type modification,
	and the blue dotted line represents the result with the $\beta$--type modification.
}
\label{fig:Mu_nq_ud}
\end{figure}

The calculated quark chemical potential dependence of the number density of $u$ and $d$ quarks in the Wigner solution is shown in Fig.~\ref{fig:Mu_nq_ud}.
It is clear that the result with the $\beta$--type modification has the largest number density.
This is physically reasonable, because the $\beta$--type modification has the smallest coupling strength in the relevant chemical potential region,
and the dressed mass is generally smaller corresponding to a smaller coupling strength, and therefore the number density of the quarks with smaller dressed mass is larger.

In the study of compact star matter, it is also necessary to take into consideration the effect of strange quarks.
The study of the solutions of DS equation has revealed that,
for a fixed coupling strength, there exists a critical current quark mass, above which the Wigner solution will disappear~\cite{LChang:2007,Williams:2007PLB,Wang:2012PRD}.
For a coupling strength fixed in vacuum, the strange quark mass is well above the critical mass,
which means that the strange quark does not have the Wigner solution for the Gaussian--type gluon model without modification,
whose coupling strength is the same as that in vacuum at any chemical potential.
In case of the gluon model with the modifications, however,
since the Wigner solution must exist at large enough chemical potential where the coupling strength is quite small,
we can then get the strange quark number density above some critical chemical potential.

\begin{figure}[!htb]
\vspace*{-3mm}
\includegraphics[width=0.45\textwidth]{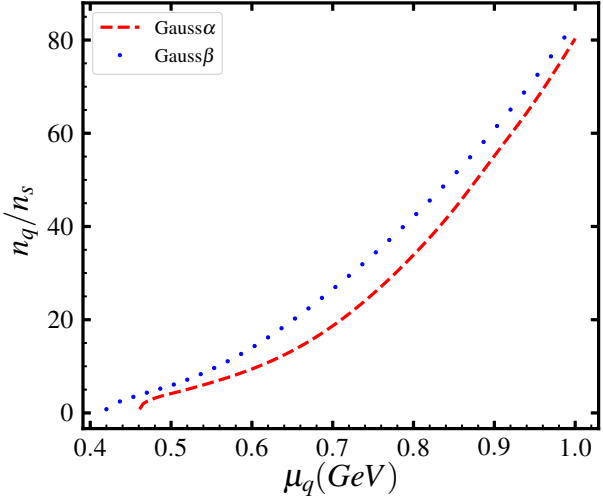}
\vspace*{-4mm}
	\caption{(color online) Calculated number density (in unit of the saturation nuclear matter density $n_{S}$) of the $s$ quark as a function of the quark chemical potential.
	The red dashed line corresponds to the result using the Gaussian-type gluon model with the $\alpha$--type modification,
	and the blue dotted line stands for the result with the $\beta$--type modification.
}
\label{fig:Mu_nq_s}
\end{figure}

The calculated number density of the the strange quarks as a function of quark chemical potential is shown in Fig.~\ref{fig:Mu_nq_s}.
One can notice from Fig.~\ref{fig:Mu_nq_s} that, in case of the gluon model with the $\alpha$--type modification,
the critical chemical potential for the strange quark to appear is $\mu_{s,c}=0.461\,$GeV,
while for the gluon model with the $\beta$--type modification, the critical chemical potential is $\mu_{s,c}=0.420\,$GeV.
The critical chemical potential in case of the $\beta$ type modification is smaller, because it has smaller coupling strength in the relevant chemical potential region.
Also, the strange quark density for the $\beta$--modification is larger, with the same reason for that the $u$ and $d$ quark number density is larger for the $\beta$--type modification.

After calculating the number density  of the quarks,
the pressure of each flavor of the quark at zero temperature can be obtained by integrating the number density:
\begin{equation}
P_{q}^{}(\mu_{q}^{}) = P_{q}^{}(\mu_{q,0}^{}) + \int_{\mu_{q,0}^{}}^{\mu_{q}^{}} \textrm{d}\mu n_{q}^{}(\mu) \, .
\end{equation}
Mathematically, the starting point of the integration $\mu_{q,0}$ can be any value,
but in our calculation, it is more convenient to choose $\mu_{u/d,0}$ to be the value of the left boundary of the coexistence region,
and $\mu_{s,0}$ to be the critical chemical potential where the Wigner solution for the $s$ quark to appear.
In Ref.~\cite{Chen:2008PRD}, the pressure difference between the Nambu and Wigner solutions is calculated by using the ``steepest-descent" approximation,
and since the Nambu solution corresponds to the vacuum at $\mu_{q,0}$,
the initial value of the integration can be chosen as $P_{q}^{}(\mu_{q,0}) = P_{W}(\mu_{q,0}) - P_{N}(\mu_{q,0})$,
	where $P_{W}$ and $P_{N}$ is the pressure of the Wigner solution, Nambu solution, respectively.
By adopting the result from Ref.~\cite{Chen:2008PRD}, we have $P_{q}^{}(\mu_{q,0})=2.58\times10^{-4}$, $4.32\times 10^{-4}$ and $4.19\times 10^{-4} \textrm{GeV}^{-4}$
in case of the Gaussian-type gluon model without modification,
   with the $\alpha$--type modification and with the $\beta$--type modification, respectively.
For the $s$ quark, we simply choose $P_{s}(\mu_{s,0})=0$ as in Ref.~\cite{Chen:2011PRD}.

The total pressure of the quark matter is the sum of the pressure of each flavor of the quarks:
\begin{equation}
P_{Q}^{}(\mu_{u}^{},\mu_{d}^{},\mu_{s}^{}) = \sum_{q=u,d,s}\tilde{P}_{q}^{}(\mu_{q}^{}) - B_{\textrm{DS}}^{} \, ,
\end{equation}
where
\begin{equation}
\tilde{P}_{q}^{}(\mu_{q}^{})\equiv \int_{\mu_{q,0}^{}}^{\mu_{q}^{}}\textrm{d}\mu n_{q}^{}(\mu) \, ,
\end{equation}
and
\begin{equation}\label{eqn:B_DS}
B_{\textrm{DS}}^{} \equiv -\sum_{q=u,d,s} P_{q}^{}(\mu_{q,0}^{}) \, .
\end{equation}

In compact star matter, the contributions from leptons is also important.
Usually, the electron and muon are considered.
The number density for the leptons is:
\begin{equation}\label{eqn:nl}
n_{l}^{} = \frac{k_{Fl}^{3}}{3\pi^{2}} \, ,
\end{equation}
where $k_{Fl}^{2} = \mu_{l}^{2} - m_{l}^{2}$ for $l=e^{-},\, \mu^{-}$.
In this work, we take $m_{e}=0.511\;$MeV and $m_{\mu^{-}}=105\;$MeV.

The quark matter in a compact star should be in $\beta$-equilibrium and electric charge neutral,
so we have:
\begin{equation}
\begin{split}
& \mu_{d}^{}= \mu_{u}^{} + \mu_{e}^{} = \mu_{s}^{} \, ,\\
& \mu_{\mu^{-}}=\mu_{e}
\end{split}
\end{equation}
\begin{equation}
\frac{2n_{u}^{} - n_{d}^{} - n_{s}^{}}{3} - n_{e}^{} - n_{\mu^{-}}^{} = 0 \, .
\end{equation}
And we have the baryon density and chemical potential as:
\begin{equation}
n_{B}^{} = \frac{1}{3}(n_{u}^{} + n_{d}^{} + n_{s}^{} ) \, ,
\end{equation}
\begin{equation}\label{muB}
\mu_{B}^{} = \mu_{u}^{} + 2\mu_{d}^{} \, .
\end{equation}

With these relations, we can calculate the properties of the $\beta$--equilibrium and charge neutral quark matter with a given baryon chemical potential (baryon density).

Except for solving the DS equation for the quark matter and the RMF for the hadron matter,
we also need to make use of some construction scheme to describe the hadron-quark phase transition.
The condition for the phase transition to occur is the chemical potential and the pressure in the two phase are equal, i.e.,
\begin{equation}
p_{H}(\mu_{B})=p_{Q}(\mu_{B}),
\end{equation}
where the footnote $H$ and $Q$ denote the hadron and quark sector,
and the pressure as a function of baryon chemical potential is determined by the hadron and quark model we implemented.
By solving the above equation, we can get the chemical potential corresponding to the hadron--quark phase transition.

The above described construction scheme is called the ``Maxwell construction",
which provides a straight forward way to describe the phase transition, and is widely used.
However, it only considers one chemical potential, the baryon chemical potential.
In the compact star matter, there are two conservation numbers, the baryon number and the charge number.
In turn, there are two chemical potentials, the baryon chemical potential and the charge chemical potential.
We should then implement the Gibbs construction to take into account both of the chemical potentials.

In Gibbs construction,
it assumes that there is a mix phase region (phase coexistence region) where both quark and hadron exist simultaneously,
and the baryon chemical potential ($\mu_{B}^{}$) and the charge chemical potential($\mu_{e}$) are the same in both the quark and the hadron phases.

In the phase coexistence region, the pressure of the two phases are the same.
And though the two phases are no longer charge neutral separately,
there will be a global charge neutral condition.
If we define the quark fraction $\chi \in [0, 1]$,
the phase transition conditions can be expressed as:
\begin{equation}\label{eqn:condition1}
p_H(\mu_B,\mu_e)=p_Q(\mu_B,\mu_e),
\end{equation}
\begin{equation}\label{eqn:condition2}
(1-\chi)n^c_H(\mu_B,\mu_e)+\chi n^c_Q(\mu_B,\mu_e)=0,
\end{equation}
where $p_{H}$, $p_{Q}$ is the pressure of the hadron, the quark phase, respectively,
which is a function of both the $\mu_{B}$ and the $\mu_{e}$.
 $n^{c}_{H}$, $n^{c}_{Q}$ is the charge density of each of the two phases,
which is determined by the corresponding hadron and quark model.

Then, combining Eqs.~(\ref{eqn:condition1}) and (\ref{eqn:condition2}),
together with the field equations of the two phases,
we can obtain the $\mu_{B}^{}$ and $\mu_{e}$ with any given quark fraction $\chi$.
By taking $\chi=0,\;1$, we can calculate the phase boundary of the coexistence region under the charge neutral and $\beta$--equilibrium condition.

\begin{figure}[!htb]
\vspace*{-3mm}
\includegraphics[width=0.45\textwidth]{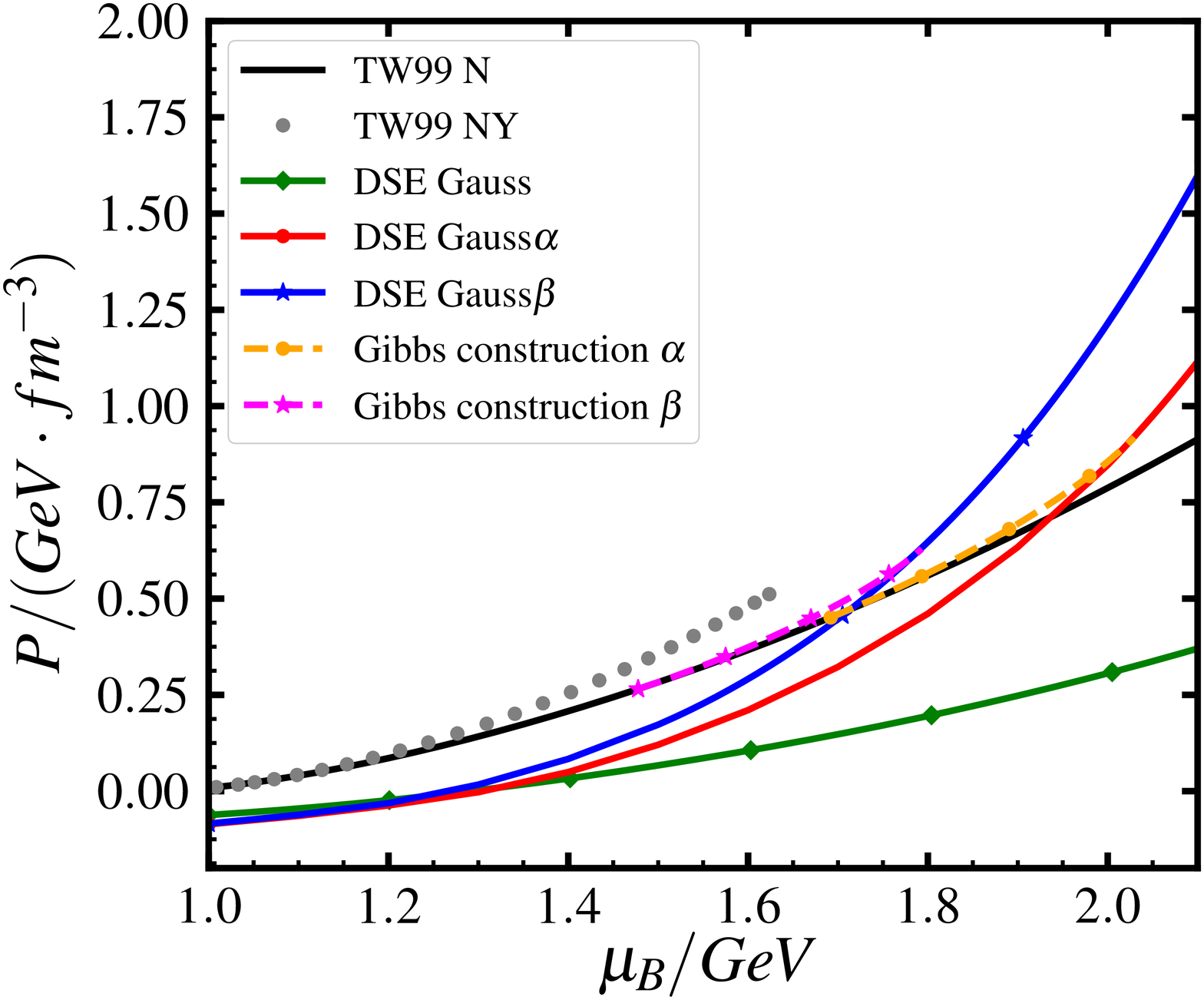}
\vspace*{-5mm}
	\caption{(color online) Calculated pressure as a function of baryon chemical potential.
	The black solid line corresponds to the pressure of hadron matter without hyperons using the RMF TW-99 model,
		the gray dotted line corresponds to that of hadron matter with hyperons in the RMF TW-99 model,
	the green line with diamond symbol corresponds to the quark matter with DS equation calculation with the Gaussian--type gluon model without modification,
	the red line with circle symbol stands for the quark matter with the DS equation calculation with the $\alpha$--type modification,
	the blue line with star symbol represents that of the the quark matter with the DS equation calculation implementing the $\beta$--type modification,
       the orange dashed line with  circle symbol corresponds to the Gibbs construction linking the RMF TW-99 hadron model and the DS equation using the $\alpha$--type quark model,
	and the pink dashed line with star symbol is that via the Gibbs construction linking the RMF TW-99 hadron model and the DS equation using the $\beta$--type quark model.
	}
\label{fig:P_Mu}
\end{figure}

The calculated baryon chemical potential dependence of the pressure of each kind matter is shown in Fig.~\ref{fig:P_Mu}.
From the figure, we can see that, among the three quark models,
the one with the $\beta$--type modification has the largest pressure,
which is in accordance with that the $\beta$--type modification has the largest quark number density.

In Fig.~\ref{fig:P_Mu}, the black solid line and gray dotted line represent the results of the calculated hadron matter pressure via the TW-99 RMF model described in the appendix.
The black solid line corresponds to the nuclear matter that only proton, neutron and leptons are considered,
	while the gray dotted line is the one with the contribution of the hyperons having been taken into consideration.
As we have mentioned, for the hadron-quark phase transition to take place,
the pressure and the chemical potential of hadron matter and those of the quark matter must be simultaneously the same, respectively.
This means that the $P$--$\mu_{B}^{}$ curve of the hadron matter and that of the quark matter must have a cross point.
However, Fig.~\ref{fig:P_Mu} manifests clearly that, in case of the Gaussian--type gluon model without any modification,
the pressure of the quark matter is too small and there is no cross point with the line of hadron matter.
This means that, for the Gaussian-type gluon model without modifications,
the coupling strength in the relevant chemical potential region is too strong and should be excluded,
 since the hadron-quark phase transition is forbidden in the model,
which is physically unlikely.
This further indicates the necessity of taking into consideration of the modification factor.
For the Gaussian--type gluon model with the $\alpha$- and $\beta$--type modification,
the phase transition chemical potential for the $\beta$--equilibrium and charge neutral matter is $\mu_{B}^{}=1.94\,$GeV, $1.71\,$GeV, respectively.
And the gray dotted line does not have cross point with all the quark lines,
	which means that the phase transition will not happen after the inclusion of hyperons.
Therefore, we will not include the hyperon effect in the following.

By implementing the Gibbs construction, we can obtain that the phase coexistence region of the $\beta$--equilibrium and global charge neutral matter is
$\mu_{B}^{} \in [1.69 , 2.02]\,$GeV,  $\mu_{B}^{} \in [1.48, 1.79]\,$GeV in case of the model with the $\alpha$--type, $\beta$--type modification, respectively, as shown in Fig.~\ref{fig:P_Mu}.

The obtained phase coexistence region and the phase boundary here is different from what we have shown in Sec.~\ref{sec::isosymmetric},
but with some simple relations:
\begin{equation}\label{eqn:iso_beta}
3\mu_{W,c}\le \mu_{G,1}\le\mu_{G,2}\le 3\mu_{N,c},
\end{equation}
where $\mu_{G,1}^{}$, $\mu_{G,2}^{}$ is the baryon chemical potential corresponding to the left, the right boundary of the coexistence region from the Gibbs construction,
$\mu_{W,c}^{}$ is the quark chemical potential corresponding to the divergence of the chiral susceptibility of the Wigner solution (DCS phase),
and $\mu_{N,c}^{}$ corresponds to the divergence of the chiral susceptibility of the Nambu solution (DCSB phase).

The first inequality in Eq.(\ref{eqn:iso_beta}) is satisfied well in our calculation,
while the last inequality is not satisfied.
This deficiency comes from both the insufficiency of the hadron model and the quark model.

By solving the DS equation of the quark(s), we are calculating the property of uniform quark matter.
However, the quarks in the hadron matter is not uniformly distributed since they are clustered as hadrons.
So the Nambu solution of DS equation is different from the real world hadron matter after the appearance of quark number density.

The RMF model also has its deficiency.
The parameters of the hadron matter model are fixed by fitting the property of the hadron matter at saturation density or lower densities
where terrestrial experiments are able to create.
Beyond the saturation density, different models give distinct results.
Still, the RMF model is more reliable than the Nambu solution of the DS equation in describing the hadron matter,
since the solution of the DS equation has not taken into account the effects of the non-uniform distribution of the quark, the surface of the cluster, and so forth.
We then conclude that the right boundary of the phase coexistence region of the matter is at $\mu_{N,c}^{} \ge {\mu_{G,2}^{}}/3 = 0.673\,$GeV,  $\mu_{N,c}^{} \ge 0.597\,$GeV for the $\alpha$--type modified, $\beta$--type modified gluon model, respectively.

\subsection{Equation of State and Mass-radius Relation of Compact Star}

In order to study the hadron--quark phase transition at zero temperature and high chemical potential (density),
one has to make use of the compact star observations to check our theoretical results since it is not able to generate such dense matter on earth.

The most important observable of compact stars is the maximum mass,
which is related directly to the EOS $P=P(\varepsilon)$ of the dense matter.

The energy density $\varepsilon$ of the dense matter under Maxwell construction is:
\begin{equation}
	\varepsilon_{\textrm{Maxwell}} = \left\{ \begin{array}{ll}
		\varepsilon_{H}, & \quad \textrm{if}\; \mu_{B}^{}<\mu_{B,c}^{},  \\[2mm]
		\varepsilon_{Q}, & \quad \textrm{if}\; \mu_{B}^{}>\mu_{B,c}^{},\\
	\end{array}\right.
\end{equation}
where $\varepsilon_{H}$ and $\varepsilon_{Q}$ is the energy density of the charge neutral hadron matter and quark matter, respectively.
$\mu_{B,c}^{}$ is the phase transition baryon chemical potential.
The pressure of the dense matter under Maxwell construction is:
\begin{equation}
	P_{\textrm{Maxwell}} = \left\{ \begin{array}{ll}
		P_{H}^{}, & \quad \textrm{if}\; \mu_{B}<\mu_{B,c}^{}, \\[2mm]
		P_{Q}^{}, & \quad \textrm{if}\; \mu_{B}>\mu_{B,c}^{} , \\
	\end{array}\right.
\end{equation}
where $P_{H}^{}$ is the pressure of the hadron matter and $P_{Q}^{}$ is the pressure of the quark matter.

For Gibbs construction, the EOS should be decomposed into three parts:
$\mu_{B}\le \mu_{G,1}$ region, $\mu_{G,1}\le \mu_{B}\le \mu_{G,2}$ region and $\mu_{B}\ge\mu_{G,2}$ region,
where $\mu_{G,1}$ and $\mu_{G,2}$ is the left and right boundary of the phase coexistence region (mixed phase).

The energy density of the mixed phase consists of the contribution of the two phases.
\begin{equation}
\varepsilon_{M} = \chi\varepsilon_{Q}(\mu_{B},\mu_{e}) + (1-\chi) \varepsilon_{H} (\mu_{B},\mu_{e}),
\end{equation}
where the footnote $M,Q$ and $H$ corresponds to the mixed, the quark, the hadron phase, respectively.
For a fixed $\mu_{B}$, $\mu_{e}$ is calculated by solving Eq.(\ref{eqn:condition1}) and Eq.(\ref{eqn:condition2}).
And the pressure for the mixed phase reads:
\begin{equation}
p_{M}^{}(\mu_{B}) = p_{H}^{} \left(\mu_{B},\mu_{e}(\mu_{B})\right) = p_{Q}^{} \left(\mu_{B},\mu_{e}(\mu_{B})\right).
\end{equation}

The energy density and the pressure under the Gibbs construction are:
\begin{equation}
	\varepsilon_{\textrm{Gibbs}} = \left\{ \begin{array}{ll}
		\varepsilon_{H} & \quad \textrm{if}\; \mu_{B}<\mu_{G,1} , \\[1mm]
             \varepsilon_{M}& \quad \textrm{if}\; \mu_{G,1}\le\mu_{B}<\mu_{G,2}, \\[1mm]
		\varepsilon_{Q} & \quad \textrm{if}\; \mu_{B}>\mu_{G,2}, \\
	\end{array}\right.
\end{equation}
\begin{equation}
	P_{\textrm{Gibbs}}^{} = \left\{ \begin{array}{ll}
		P_{H}^{} & \quad \textrm{if}\; \mu_{B}<\mu_{G,1}, \\[1mm]
            P_{M}^{} &\quad \textrm{if}\; \mu_{G,1}\le\mu_{B}<\mu_{G,2}, \\[1mm]
		P_{Q}^{} & \quad \textrm{if}\; \mu_{B}>\mu_{G,2} . \\
	\end{array}\right.
\end{equation}

\begin{figure}[!htb]
\vspace*{-3mm}
\includegraphics[width=0.45\textwidth]{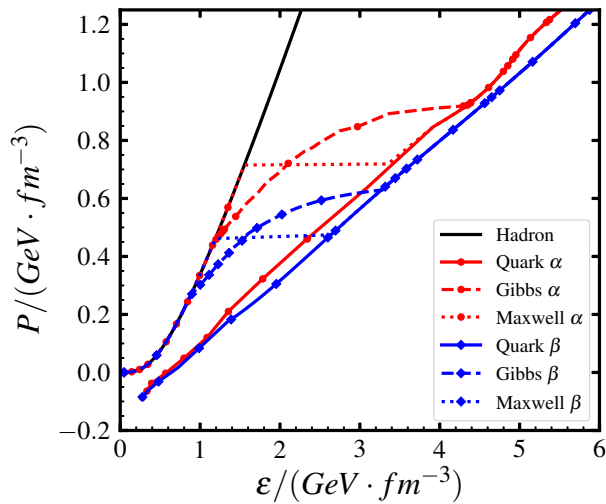}
\vspace*{-4mm}
\caption{(color online) Calculated equation of state.
	The black solid line is the result of the pure hadron matter via the TW-99 RMF model,
      the red lines with circle symbol stand for the results using the $\alpha$--type modification,
	the blue lines with diamond symbol correspond the result implementing the $\beta$--type modification.
	The solid lines with symbols represent the results of the pure quark matter,
the dashed lines with symbols correspond to the results via the Gibbs construction,
and the dotted lines with symbols stand for the results with the Maxwell construction.}
\label{fig:EoS}
\end{figure}

The calculated EOSs of the pure hadron, pure quark and hybrid (mixed phase) matter are illustrated in Fig.~\ref{fig:EoS}.
As we can recognize  from Fig.~\ref{fig:EoS},
the pure quark matter has a softer EOS than the pure hadron matter,
and both the two different construction schemes connect the EOSs of the hadron matter and the quark matter.
In case of the Maxwell construction, there is a density region where the pressure is constant,
which corresponds to a phase transition with a constant chemical potential.
However, in compact stars, the pressure must have a gradient in order to resist its own gravity,
so that such a constant pressure region will not appear inside compact star.
The phase transition under the Maxwell construction corresponds to a sudden change in the energy density between the quark core and the hadron shell.

The mass--radius relation of a compact star can be calculated
by solving the Tolman-Oppenheimer-Volkov (TOV) equation:
\begin{equation}
\frac{\textrm{d}P}{\textrm{d}r}=-\frac{G}{r^2}(m(r)+4\pi Pr^3)(\epsilon+P)(1-2Gm(r)/r)^{-1},
\end{equation}
where $G$ is gravitational constant and $m(r)$ is the mass inside a radius $r$:
\begin{equation}
m(r)=\int_{0}^{r} 4\pi R^{2} \varepsilon \textrm{d}R.
\end{equation}
Then given the EOS as input,
and with a given center density,
we can integrate the TOV equation from inside out to get the mass and radius of the compact star.

For the pure hadron star and the hybrid star, at small density region, we make use of the Baym--Pethick--Sutherland (BPS) EOS~\cite{Baym:1971APJ}.
For the pure quark star, we integrate to the surface where the pressure is zero.

\begin{figure}[!htb]
\vspace*{-3mm}
\includegraphics[width=0.45\textwidth]{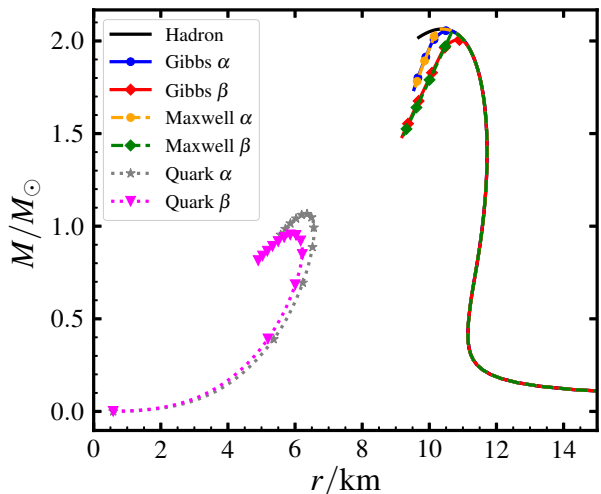}
\vspace*{-5mm}
\caption{(color online) Calculated mass-radius relation of the compact star in the cases we take consideration.
	The black solid line corresponds to the pure hadron star,
the solid lines with symbols stand for the results via the Gibbs construction,
the dashed lines with symbols represent the results with the Maxwell construction,
and the dotted lines with symbols correspond to the results of the pure quark matter.
	The $\alpha$, $\beta$ in the legend denotes the DS equation result by implementing the $\alpha$-, $\beta$--type modification, respectively.}
\label{fig:Mass_Radius}
\end{figure}

The calculated mass-radius relation of the compact star in the cases of our consideration is shown in Fig.~\ref{fig:Mass_Radius}.
From Fig.~\ref{fig:Mass_Radius}, one can notice apparently that the pure quark star has a much smaller maximum mass than the pure hadron star and hybrid star,
	 and the maximum mass is $1.07M_{\odot}$, $0.95M_{\odot}$ for the $\alpha$-, $\beta$--type modification, respectively.

For pure hadron star, the maximum mass is $2.06M_{\odot}$.
From Fig.~\ref{fig:Mass_Radius}, we can see that at large radius and small mass,
 the mass-radius curves for hybrid star and pure hadron star are exactly the same,
 because for hybrid star with relatively small mass, the inner density is not large enough for the quark matter to appear.

When the mass of the hybrid star is large enough, the quark matter begins to appear in the composing matter of the star,
     and the mass--radius curves begin to deviate from the pure hadron one.
As can be seen from Fig.~\ref{fig:Mass_Radius}, for both $\alpha$ and $\beta$ modification, and for both Gibbs and Maxwell construction,
   the deviation begin to appear at over $\sim 2M_{\odot}$,
   which means that the hadron-quark phase transition happens at very high density and the mass--radius relation is determined mainly by the EOS of the hadron sector.
If the quark sector is described by DS equation with the $\alpha$--type modification, the maximum mass will be $2.06M_{\odot}$ for both Gibbs and Maxwell construction.
If the quark sector is described by DS equation with the $\beta$--type modification,
   the maximum will be $2.00M_{\odot}$ for Gibbs construction and $2.05M_{\odot}$ for Maxwell construction.

\section{Summary}
\label{Sec::summary}
The Dyson-Schwinger equation has been implemented to study the phase transition and the phase structure of QCD matter.
However, the possible dependence of coupling strength on the quark chemical potential is not well studied.
In this paper, we take into account several types of the chemical potential dependence,
and make use of the nuclear liquid-gas phase transition to fix the parameter(s).

For the modification factor we propose, the boundary of the phase coexistence region of the iso-symmetric matter is reduced than that without modification.
The phase coexistence region is $\mu_{q} \in [0.322 , 0.500]\,$GeV for the Gaussian--type gluon model without any modification,
    and becomes $\mu_{q} \in [0.271, 0.309]\,$GeV, $[0.276, 0.308]\,$GeV in case of the $\alpha$-, $\beta$--type modification, respectively.

However, by taking into account the $\beta$--equilibrium and charge neutral condition for the matter inside compact stars,
and comparing the results from the DS equation calculation for the quark sector and the RMF hadron model,
    we find that the quark pressure is too low for the Gaussian--type gluon model without any modification,
    and the hadron--quark phase transition is prohibited, which is physically unlikely.
This means that the adoption of modification is necessary.

By implementing the RMF model for hadron matter and the DS equation calculation for the quark sector,
   we find that the phase transition chemical potential under the Maxwell construction is $\mu_{B}=1.94$, $1.71\,$GeV for the $\alpha$-, $\beta$--type modification, respectively.
The phase coexistence region under the Gibbs construction is $\mu_{B} \in [1.69, 2.02]\,$GeV, $[1.48, 1.79]\,$GeV in case of the $\alpha$-, $\beta$--type modification, respectively.
The upper boundary of this mixed phase region is not compatible with the iso-symmetric result,
    and discrepancy comes from the uncertainty of both the DS equation approach and the RMF model in the phase coexistence region.
Since the RMF model is more accurate in describing the hadron phase, we conclude that this boundary indicates that $\mu_{N,c}\ge 0.673\,$GeV, $0.579\,$GeV,
for the $\alpha$- $\beta$--type modification, respectively,
	where $\mu_{N,c}^{}$ is the quark chemical potential at which the Nambu solution (DCSB phase) disappears.

We also obtain the mass-radius relation for compact stars in the cases we considered.
The maximum mass of the pure hadron star is $2.06M_{\odot}$.
If the quark sector is described using $\alpha$--type modification, the maximum mass of the hybrid star is the same as that of pure hadron star.
If the quark sector is described by implementing $\beta$--type modification,
   the maximum mass of the hybrid star is $2.00M_{\odot}$ and $2.05M_{\odot}$ for the Gibbs construction, the Maxwell construction, respectively.
Therefore, for hybrid stars, the maximum mass is almost the same since the phase transition happens at very high density.
However, the maximum mass of the pure quark star is only about $1M_{\odot}$.

In this work, the possible appearance of hyperons in the dense star matter is not considered,
   because it prevents from the happening of the hadron--quark phase transition.
This problem can be fixed by implementing the 3-window construction method instead of the Gibbs and Maxwell constructions,
as has been done in our previous work~\cite{Bai:2018PRD}.

Although we have made use of the liquid-gas phase transition chemical potential to fix the parameters of the modification function in our DS equation claculation,
		 it is still impossible to recover all the properties of the hadron matter from first--principle calculation.
For example, there should be a density gap at liquid-gas phase transition.
Also, the saturation density of nuclear matter cannot be recovered by the Nambu solution of the DS equation.
To fix these problems, taking into account the dressing effect of the quark--gluon interaction vertex or/and the interface effect between quarks and hadrons (clustered quarks)  in the DS equation approach should be an efficient way.
The related work is under progress.

\begin{acknowledgments}
The work was supported by the National Natural Science Foundation of China under Contracts No.\ 11435001 and
No. 11775041 .
%
%
\end{acknowledgments}

\begin{appendix}

\section{Relativistic Mean Field Theory}

The relativistic mean field (RMF) theory~\cite{Walecka:1974AP,Boguta:1977,Serot:1986ANP} is a successful
theory which describes the property of the hadron matter in the density region relevant in compact stars~\cite{Glendenning:2000,Oertel:2017RMP}.
There are many different RMF models,
with different parameterizations which are calibrated by the properties of dense nuclear matter.
However, most of the models satisfy only a small number of experimental data.
In Ref.~\cite{Dutra:2014PRC}, hundreds of RMF models are taken to fit the experimental data,
and the TW-99 model~\cite{Typel:1999NPA} is found to be one of the best model
that can reproduce most of the nuclear matter properties.
Also, TW-99 model generates an EOS that is stiff enough to support a 2-solar-mass neutron star~\cite{Dutra:2016PRC},
and we have taken such a model to construct a massive hybrid star in our previous work~\cite{Bai:2018PRD}.

The Lagrangian of the TW-99 model for the hadron matter is
\begin{equation}\label{eqn:Lagrangian}
\mathcal{L}=\mathcal{L}_{B}^{} + \mathcal{L}_{\textrm{lep}}^{} + \mathcal{L}_{M}^{} + \mathcal{L}_{\textrm{int}}^{} \, ,
\end{equation}
where $\mathcal{L}_{B}^{}$ is the Lagrangian of free baryons, which reads
%
%
\begin{equation}
\mathcal{L}_{\textrm{B}}^{} = \sum_{\textrm{i}}{\bar{\Psi}_{i}^{}} (i\gamma_{\mu}^{} \partial^{\mu} - m_{i}^{}){\Psi_{i}^{}}\, ,
\end{equation}
where $i=p,n$ stands for proton and neutron.
If we want to take into consideration the effect of hyperons,
   we can take $i=p,\,n,\,\Lambda,\,\Sigma^{\pm,0},\,\Xi^{-,0}$ for the baryon octect.

$\mathcal{L}_M$ in Eq.(\ref{eqn:Lagrangian}) is the Lagrangian of mesons,
\begin{equation}
\begin{split}
\mathcal{L}_{M}^{} =&  \frac{1}{2}\left(\partial_{\mu}^{} \sigma \partial^{\mu} \sigma - m_{\sigma}^{2} \sigma^{2} \right) -\frac{1}{4} \omega_{\mu\nu}^{} \omega^{\mu\nu}
- \frac{1}{2}m_{\omega}^{2} \omega_{\mu}\omega^{\mu} \\
& -\frac{1}{4}\boldsymbol{\rho}_{\mu\nu}^{} \boldsymbol{\rho}^{\mu\nu} -\frac{1}{2}m_{\rho}^{2} \boldsymbol{\rho}_{\mu} \boldsymbol{\rho}^{\mu} \, ,
 \end{split}
\end{equation}
where $\sigma$, $\omega_{\mu}^{}$, and $\boldsymbol{\rho}_{\mu}^{}$ are the
isoscalar-scalar, isoscalar-vector and isovector-vector meson field, respectively.
$\omega_{\mu\nu}^{}=\partial_{\mu}\omega_{\nu} - \partial_{\nu}\omega_{\mu}$,
$\boldsymbol{\rho}_{\mu\nu}^{} =\partial_{\mu}\boldsymbol{\rho}_{\nu} - \partial_{\nu}\boldsymbol{\rho}_{\mu}\,$.

The $\mathcal{L}_{\textrm{int}}^{}$ in Eq. (\ref{eqn:Lagrangian}) is the Lagrangian
 describing the interactions between baryons which are realized by exchanging the mesons:
\begin{equation}
\begin{split}
\mathcal{L}_{\textrm{int}}^{} = & \sum_{B} g_{\sigma B}^{} {\bar{\Psi}_{B}^{}} \sigma {\Psi_{B}^{}}
-g_{\omega B}^{} {\bar{\Psi}_{B}^{}} \gamma_{\mu}^{} \omega^{\mu} {\Psi_{B}^{}}\\
		&- g_{\rho B}^{} {\bar{\Psi}_{B}^{}} \gamma_{\mu}^{} \boldsymbol{\tau}_{B}^{} \cdot \boldsymbol{\rho}^{\mu} {\Psi_{B}^{}} \, ,
	\end{split}
\end{equation}
where $g_{iB}^{}$ for $i=\sigma$, $\omega$, $\rho$ are the coupling strength parameters between baryons and mesons, which depend on the baryon density.

In some other literatures, the self-interaction of $\sigma$-meson,
the cross interaction between different kind mesons and the effect of the isovector-scalar $\delta$-meson are included explicitly
(see, e.g., the review in Refs.~\cite{Oertel:2017RMP,Dutra:2014PRC}).
In TW-99 parameterization, however, all these terms are taken zero,
and their effects are represented in the density dependence of the coupling constants.
For nucleons, the coupling constants are
\begin{equation}
{g_{iN}^{}}(n_{B}^{}) = {g_{iN}^{}} (n_{s}) f_{i}^{} (x),  \qquad \quad \textrm{for} \quad i=\sigma,\omega, \rho,
\end{equation}
where $n_{B}^{}$ is the baryon density, $n_{s}$ is the saturation nuclear matter density  and
$x= {n_{B}^{}}/{n_{s}^{}}$.
The density function can be chosen as~\cite{Typel:1999NPA}:
\begin{equation}
f_{i}^{}(x)  = a_{i}^{} \frac{1+b_{i}^{} (x + d_{i}^{})^{2}}{1 + c_{i}^{}(x+d_{i}^{})^{2}}, \qquad \textrm{for} \qquad i=\sigma,\omega \, ,
\end{equation}
\begin{equation}
f_{\rho}^{}(x)  = \textrm{exp}\left[-a_{\rho}^{} (x-1) \right] \, ,
\end{equation}
where the parameters $a_{i}^{}$, $b_{i}^{}$, $c_{i}^{}$, $d_{i}^{}$ and $g_{iN}^{}(\rho_{\textrm{sat}}^{})$
are fixed by fitting the properties of the nuclear matter at the saturation density,
and their values are listed in Table \ref{tab:gi}.

\begin{table}[htb]
\caption{Parameters of the mesons and their couplings (taken from Ref.~\cite{Typel:1999NPA}).}\label{tab:gi}
\begin{tabular}{cccc}
\hline
Meson i & $\sigma$ & $\omega$ & $\rho$ \\
\hline
~$m_i$(MeV)~ & 550 & 783 & 763\\
$g_{iN}^{}(n_{s}^{})$ & ~$10.72854$~ & ~$13.29015$~ & ~$7.32196$~ \\
$a_{i}^{}$ & ~$1.365469$~ & ~$1.402488$~ & ~$0.515$~ \\
$b_{i}^{}$ & ~$0.226061$~ & ~$0.172577$~ & {} \\
$c_{i}^{}$ & ~$0.409704$~ & ~$0.344293$~ & {} \\
$d_{i}^{}$ & ~$0.901995$~ & ~$0.983955$~ & {} \\
\hline
\end{tabular}
\end{table}

For hyperons, we represent them with the relation between the hyperon coupling and the nucleon coupling as:
$\chi_{\sigma}^{}=\frac{g_{\sigma Y}^{}}{g_{\sigma N}^{}}$,
$\chi_{\omega}^{}=\frac{g_{\omega Y}^{}}{g_{\omega N}^{}}$,
$\chi_{\rho}^{}=\frac{g_{\rho Y}^{}}{g_{\rho N}^{}}$.
On the basis of hypernuclei experimental data, we choose them as those in Refs.~\cite{Glendenning:2000,Dutra:2016PRC}:
$\chi_{\sigma}^{}=0.7$, $\chi_{\omega}^{}=\chi_{\rho}^{}=0.783$.

The $\mathcal{L}_{\textrm{lep}}^{}$ is the Lagrangian for leptons, which are treated as free fermions:
\begin{equation}
\mathcal{L}_{\textrm{lep}}^{}=\sum_{l}^{} {\bar{\Psi}_{l}^{}}(i\gamma_{\mu}^{} \partial^{\mu} - m_{l}^{}) {\Psi_{l}^{}} \, ,
\end{equation}
and we include only the electron and muon in this paper.

The field equations can be derived by differentiating the Lagrangian.
Under RMF approximation, the system is assumed to be in the static, uniform ground state.
The partial derivatives of the mesons all vanish,
only the 0-component of the vector meson and the 3rd-component of the isovector meson survive and can be replaced with the corresponding expectation values.
The field equations of the mesons are then:
\begin{equation}\label{eqn:sigma}
m_{\sigma}^{2} \sigma =\sum_{B} g_{\sigma B}^{} \langle {\bar{\Psi}_{B}^{}} {\Psi_{B}^{}} \rangle \, ,
\end{equation}
\begin{equation}\label{eqn:omega}
m_{\omega}^{2} \omega_{0}^{} = \sum_{B} g_{\omega B}^{} \langle {\bar{\Psi}_{B}^{}} \gamma_{0}^{} {\Psi_{B}^{}} \rangle \, ,
\end{equation}
\begin{equation}\label{eqn:rho}
m_{\rho}^{2} \rho_{03}^{} = \sum_{B} g_{\rho B}^{} \langle {\bar{\Psi}_{B}^{}} \gamma_{0}^{} \tau_{3B}^{} {\Psi_{B}^{}} \rangle \, ,
\end{equation}
where $\tau_{3B}^{}$ is the 3rd-component of the isospin of baryon $B$.

The equation of motion (EoM) of the baryon is:
\begin{equation}\label{EOM}
\left[\gamma^{\mu} (i\partial_{\mu}^{} - \Sigma_{\mu}^{}) - (m_{B}^{} - g_{\sigma B}^{} \sigma) \right] {\Psi_{B}^{}} = 0 \, ,
\end{equation}
where
\begin{equation}
\Sigma_{\mu}^{} = g_{\omega B}^{} \omega_{\mu} + g_{\rho B}^{} \boldsymbol{\tau}_{B}^{} \cdot \boldsymbol{\rho}_{\mu}^{} + \Sigma_{\mu}^{\textrm{R}}.
\end{equation}
The $\Sigma_{\mu}^{\textrm{R}}$ is called the ``rearranging" term,
which appears because of the density-dependence of the coupling constant, and reads
\begin{equation}\label{eqn:rearrange}
\begin{split}
\Sigma_{\mu}^{\textrm{R}} =&  \frac{j_{\mu}^{}}{n_{B}} \bigg(\frac{\partial g_{\omega B}^{}} {\partial n_{B}}\bar{\Psi}_{B}^{} \gamma^{\nu} \Psi_{B}^{} \omega_{\nu}^{}\\
	 &+ \frac{\partial g_{\rho B}^{}}{\partial n_{B}}\bar{\Psi}_{B}^{} \gamma^{\nu} \boldsymbol{\tau}_{B}^{} \cdot \boldsymbol{\rho}_{\nu}^{} \Psi_{B}^{}
		-\frac{\partial g_{\sigma B}^{}}{\partial n_{B}}\bar{\Psi}_{B}^{} \Psi_{B}^{} \sigma \bigg) \, ,
	\end{split}
\end{equation}
where $j_{\mu}^{} = \bar{\Psi}_{B}^{} \gamma_{\mu} \Psi_{B}^{}$ is the baryon current.

Under the EoM of Eq. (\ref{EOM}),
the baryons behave as quasi-particles with effective mass
\begin{equation}\label{eqn:mstar}
m_{B}^{\ast} = m_{B}^{} - g_{\sigma B}^{} \sigma \, ,
\end{equation}
and effective chemical potential:
\begin{equation}\label{eqn:mustar}
\mu_{B}^{\ast} = \mu_{B}^{} - g_{\omega B}^{} \omega_{0}^{} - g_{\rho B}^{} \tau_{3B}^{} \rho_{03}^{} -\Sigma_{\mu}^{\textrm{R}} \, .
\end{equation}

One can then get the baryon (number) density:
\begin{equation}\label{eqn:ni}
n_{B}^{} \equiv \langle {\bar{\Psi}_{B}^{}} \gamma^{0} {\Psi_{B}^{}} \rangle =\gamma_{B}^{}\int\frac{\textrm{d}^3k}{(2\pi)^3} = \gamma_{B}^{}\frac{k_{FB}^{3}}{6\pi^{2}} \, ,
\end{equation}
where $k_{FB}^{} =\sqrt{\mu_{B}^{\ast 2} - m_{B}^{\ast 2}}$ is the Fermi momentum of the particle,
$\gamma_B=2$ is the spin degeneracy.
And the scalar density is:
\begin{equation}\label{eqn:nis}
\begin{split}
	\rho_{B}^{s} \equiv  \langle {\bar{\Psi}_{B}^{}} {\Psi_{B}^{}} \rangle &= \gamma_{B}^{}\int\frac{\textrm{d}^3k}{(2\pi)^{3}}\frac{m_{B}^{\ast}}{\sqrt{k^{2} + m_{B}^{\ast 2}}}\\
	&=\gamma_{B}^{}\frac{m_{B}^{\ast}}{4\pi^{2}}\bigg[k_{FB}^{} \mu_{i}^{\ast} - m_{B}^{\ast 2} \textrm{ln}\bigg(\frac{k_{FB}^{} + \mu_{B}^{\ast}}{m_{B}^{\ast}}\bigg)\bigg] \, .
	\end{split}
\end{equation}

The density of a kind of leptons is the same as that for baryons,
except that the effective mass and the effective chemical potential should be replaced with the corresponding mass and chemical potential of the lepton.

The matter in the star composed of hadrons should be in $\beta$-equilibrium.
Since there are two conservative charge numbers: the baryon number and the electric charge number,
all the chemical potential can be expressed with the baryon chemical potential and the electron chemical potential:
\begin{equation}\label{eqn:beta}
\mu_{i}^{} = B \mu_{B}^{} - Q \mu_{e}^{} \, ,
\end{equation}
where $B$ and $Q$ is the baryon number, electric charge number for the particle $i$, respectively.
Since neutron has one baryon number and zero charge number, we can take $\mu_{B}=\mu_{n}$ where $\mu_{n}$ is the neutron chemical potential.

Then, combining Eqs.~(\ref{eqn:sigma}), (\ref{eqn:omega}), (\ref{eqn:rho}), (\ref{eqn:rearrange}), (\ref{eqn:mstar}),
(\ref{eqn:mustar}), (\ref{eqn:ni}), (\ref{eqn:nis}), (\ref{eqn:nl}) and (\ref{eqn:beta}),
together with the charge neutral condition:
\begin{equation}
n_{p}^{} + n_{\Sigma^{+}}^{}  = n_{e}^{} + n_{\mu^{-}}^{} + n_{\Sigma^{-}}^{}
+n_{\Xi^{-}}^{} \, ,
\end{equation}
one can determine the ingredients and the properties of the hadron matter with any given baryon density $n_{B}^{}\,$.

The EOS of the hadron matter can be calculated from the energy-momentum tensor:
\begin{equation}
T^{\mu\nu} = \sum_{\phi_i} \frac{\partial\mathcal{L}}{\partial(\partial_{\mu}\phi_{i}^{})}\partial^{\nu}\phi_{i}^{} - g^{\mu\nu}\mathcal{L}.
\end{equation}
The energy density $\varepsilon$ is:
\begin{equation}
\begin{split}
\varepsilon=\langle T^{00}\rangle =\sum_{i=B,l}\varepsilon_{i}^{} +\frac{1}{2}m_{\sigma}^{2} \sigma^{2} +\frac{1}{2}m_{\omega}^{2} \omega_{0}^{2} + \frac{1}{2}m_{\rho}^{2} \rho_{03}^{2} \, ,
	\end{split}
\end{equation}
where the contribution of the baryon $B$ to the energy density is:
\begin{equation}
\begin{split}
\varepsilon_{B}^{}  &= \gamma_{B}^{}\int\frac{\textrm{d}^3k}{(2\pi)^3}\sqrt{k^{2} +m_{B}^{\ast 2}}\\
	&= \! \gamma_{B}^{}\frac{1}{4\pi^{2}} \Big[ 2\mu_{B}^{\ast 3} k_{FB}^{} \! - \! m_{B}^{\ast 2} \mu_{B}^{\ast} k_{FB}^{}
\! - \! m_{B}^{\ast 4}\textrm{ln}\Big(\! \frac{\mu_{B}^{\ast} \! + \! k_{FB}^{}}{m_{B}^{\ast}} \! \Big) \Big].
\end{split}
\end{equation}

The contribution of the leptons to the energy density can be written in the similar form as baryons with a spin degeneracy parameter $\gamma_{l}^{} = 2 $,
except that the effective mass and effective chemical potential should be replaced with those of the leptons, respectively.

As for the pressure of the system, we can determine that with the general formula:
\begin{equation}
P =\sum_{i} \mu_{i}^{} \rho_{i}^{} - \varepsilon \, .
\end{equation}

\end{appendix}

\end{document}